\documentclass[12pt, preprint]{aastex}
\usepackage{url}
\usepackage{hyperref}

\shorttitle{2MASS 1207 b}
\shortauthors{Skemer et al.}

\begin{document}

\title{Evidence Against an Edge-On Disk Around the Extrasolar Planet, 2MASS 1207 b and a New Thick Cloud Explanation for its Under-Luminosity\footnote{The observations reported here were obtained at the Gemini Observatory, which is operated by the Association of Universities for Research in Astronomy, Inc., under a cooperative agreement with the NSF on behalf of the Gemini partnership: the National Science Foundation (United States), the Science and Technology Facilities Council (United Kingdom), the National Research Council (Canada), CONICYT (Chile), the Australian Research Council (Australia), CNPq (Brazil), and CONICET (Argentina).} \footnote{Based on observations collected at the European Southern Observatory, Chile (ESO Programmes 073.C-0469, 274.C-5036, 274.C-5057 and 083.C-0283).} \footnote{Based on observations collected with the SMARTS 1.3 meter, operated by the SMARTS consortium.}}

\author{Andrew J. Skemer$^{1}$, Laird M. Close$^{1}$, L\'{a}szl\'{o} Sz\H{u}cs$^{2,3}$, D\'{a}niel Apai$^{3}$, Ilaria Pascucci$^{3}$ and Beth A. Biller$^{4}$}
\affil{$^{1}$Steward Observatory, Department of Astronomy, University of Arizona, 933 N. Cherry Ave, Tucson, AZ 85721}
\affil{$^{2}$Department of Experimental Physics, University of Szeged, Szeged, D\'{o}m t\'{e}r 9, 6720 Hungary}
\affil{$^{3}$Space Telescope Science Institute, 3700 San Martin Drive, Baltimore, MD 21218}
\affil{$^{4}$Max-Planck-Institut fur Astronomie, Knigstuhl 17, D-69117 Heidelberg, Germany}

\begin{abstract}
Since the discovery of the first directly-imaged, planetary-mass object, 2MASS 1207 b, several works have sought to explain a disparity between its observed temperature and luminosity.  Given its known age, distance, and spectral type, 2MASS 1207 b is under-luminous by a factor of $\sim$10 ($\sim$2.5 mags) when compared to standard models of brown-dwarf/giant-planet evolution.  In this paper, we study three possible sources of 2MASS 1207 b's under-luminosity.  First, we investigate \citet{2007ApJ...657.1064M}'s hypothesis that a near edge-on disk, comprising large, gray-extincting grains, might be responsible for 2MASS 1207 b's under-luminosity.  After radiative transfer modeling we conclude that the hypothesis is unlikely due to the lack of variability seen in multi-epoch photometry and unnecessary due to the increasing sample of under-luminous brown-dwarfs/giant-exoplanets that cannot be explained by an edge-on disk.  Next, we test the analogous possibility that a spherical shell of dust, could explain 2MASS 1207 b's under-luminosity.  Models containing enough dust to create $\sim$2.5 mags of extinction, placed at reasonable radii, are ruled out by our new Gemini/T-ReCS 8.7$\micron$ photometric upper-limit for 2MASS 1207 b.  Finally, we investigate the possibility that 2MASS 1207 b is intrinsically cooler than the commonly used AMES-DUSTY fits to its spectrum, and thus it is not, in fact, under-luminous. New, thick cloud model grids by \citet{2011arXiv1102.5089M} fit 2MASS 1207 b's 1-10$\micron$ SED well, but they do not quite fit its near-infrared spectrum.  However, we suggest that with some ``tuning", they might be capable of simultaneously reproducing 2MASS 1207 b's spectral shape and luminosity.  In this case, the whole class of young, under-luminous brown-dwarfs/giant-exoplanets might be explained by atmospheres that are able to suspend thick, dusty clouds in their photospheres at lower temperatures than field brown-dwarfs.
\end{abstract}

\section{Introduction}
Until recently, the vast majority of our knowledge regarding exoplanets has come from radial-velocity studies.  Now with ground and space-based dedicated instruments, transit studies are discovering and characterizing the atmospheres of hot jupiters/neptunes at a rapid pace.  The frontier for exoplanet studies is the direct detection and characterization of exoplanets, using high-contrast imagers/spectrographs.  The emphasis in the direct detection field, thus far, has been on the discovery of planets.  Now with a host of directly-detected planetary mass objects and candidates, we can proceed with multi-wavelength characterizations of their atmospheres, the primary purpose of which, at least initially, is to determine the relationships between their masses, ages, radii, and spectra.

In this paper we concentrate on one object, 2MASS 1207 b, the first directly detected, planetary-mass companion, which was found by \citet{2004AA...425L..29C} 0.78" from the brown dwarf 2MASSWJ 1207334-393254 (hereafter 2MASS 1207 A).  2MASS 1207 A and its companion, 2MASS 1207 b, were verified to be co-moving by \citet{2005AA...438L..25C} and \citet{2006ApJ...652..724S}.  Since 2MASS 1207 A is known to be a member of the young TW Hya group \citep{2002ApJ...575..484G}, \citet{2004AA...425L..29C} assign 2MASS 1207 A and b an age of $8\pm^{4}_{3}$ Myr.  Based on their measured JHKsL' fluxes, \citet{2004AA...425L..29C}, estimate 2MASS 1207 b's mass to be 5$\pm$2 $M_{\rm jup}$ with an effective temperature of 1250$\pm$200 K, assuming a distance of 70 pc.

Several groups have refined the distance of 2MASS 1207 with the moving cluster method \citep{2005ApJ...634.1385M,2007ApJ...668L.175M} and with trigonometric parallax measurements \citep{2007ApJ...669L..41B,2007ApJ...669L..45G,2008AA...477L...1D}.  For the remainder of this paper, we adopt the weighted average of the parallax measurements (52.8$\pm$1.0 pc) as the distance to 2MASS 1207.  At this distance, the inferred luminosity of 2MASS 1207 b is considerably lower ($\sim$2.5 mags) than models predict given its age and spectral-type temperature.\footnote{For the remainder of this paper, we use the term ``under-luminous" to describe this phenomenon.  Other authors \citep[see for example][]{2006ApJ...651.1166M} invert the problem using $L=\frac{4 \pi R^{2}}{D^{2}} \sigma T_{\rm eff}^{4}$, with a fixed (known) $R$, and describe $T_{\rm eff}$ as ``unusually cool" for its spectral type.}

The source of this under-luminosity is examined in great detail by \citet{2007ApJ...657.1064M}.  With low-resolution H/K spectra, and the use of accurate distance measurements, \citet{2007ApJ...657.1064M} find the temperature of 2MASS 1207 b to be 1600$\pm$100 K \citep[using AMES-DUSTY models;][]{2001ApJ...556..357A} and the bolometric luminosity to be -4.72$\pm$0.14 ($log\frac{L}{L_{\sun}}$).  This luminosity is a factor of $\sim$10 ($\sim$2.5 mags) less than the values predicted by AMES-DUSTY models.  To explain their results, \citet{2007ApJ...657.1064M} invoke a ``gray-extinction", edge-on disk model after arguing against distance/age errors, model errors, and interstellar extinction.  A highly inclined disk around 2MASS 1207 b is, \textit{a priori}, highly improbable.  However, 2MASS 1207 A is known to have a disk at a high ($>$60$^{\circ}$) inclination \citep{2004AA...427..245S,2005ApJ...629L..41S}, and disks around young binary stars tend to be close to aligned \citep{2004ApJ...600..789J,2006AA...446..201M}.  Perhaps the best evidence for the edge-on disk model is that other alternatives presented by \citet{2007ApJ...657.1064M} and \citet{2007ApJ...668L.175M} are even more unlikely.  

The source of 2MASS 1207 b's apparent under-luminosity is important in the context of other low-mass objects, some of which (HD 203030 B, HN Peg B, and HR 8799 bcde) also appear to be less luminous than their temperatures suggest, while others (GQ Lup b, 2MASS 1207 A, AB Dor C and Beta Pic b) appear normal \citep[][Bonnefoy et al. in prep]{2006ApJ...651.1166M,2007ApJ...654..570L,2008Sci...322.1348M,2010ApJ...723..850B,2010arXiv1011.4918M,2007ApJ...656..505M,2007ApJ...657.1064M,2007ApJ...665..736C}.  If 2MASS 1207 b's under-luminosity is indeed explained by an edge-on disk, other under-luminous low-mass objects might, at least individually, have their own reasons for being abnormal (such as inaccurate distance or age measurements).  However, if 2MASS 1207 b's under-luminosity is not the result of an edge-on disk, it is very difficult to explain how an age/distance inaccuracy could explain its temperature/luminosity relationship, given its TWA cluster membership, established trigonometric parallax, and 2MASS 1207 A's normal temperature/luminosity relationship.  Thus, alternative hypotheses should be considered that can explain 2MASS 1207 b's under-luminosity in the context of other under-luminous brown-dwarfs/giant-planets.

After presenting new observations and previously published data (Section \ref{Observations}), we investigate the source of 2MASS 1207 b's under-luminosity, in three parts, beginning with a test of the edge-on disk hypothesis (Section \ref{Modeling}).  We first construct RADMC radiative transfer models of a hypothetical disk around 2MASS 1207 b.  These models contribute to our discussion of the expected variability of an edge-on disk, compared to observations of 2MASS 1207 b.  We also investigate the likelihood that a system like 2MASS 1207 b would be oriented on the sky nearly edge-on.  Finally, we call attention to several other under-luminous brown-dwarf/giant-exoplanets that are difficult to explain with near edge-on disks.  Analogous to the edge-on disk hypothesis, is the idea that an isotropic shell of dust could cause the same gray-extinction as an edge-on disk, without the geometric constraints (Section \ref{Modeling DUSTY}).  We model this situation with the DUSTY radiative transfer code, and evaluate the models in the context of new, deep 8.7$\micron$ images from Gemini/T-ReCS, which provide a meaningful upper-limit on the mid-infrared brightness of 2MASS 1207 b for the first time.  After studying the edge-on disk and dust shell hypotheses, both of which assume that 2MASS 1207 b is a $T_{\rm eff}=1600$K brown-dwarf/giant-exoplanet with a low luminosity explained by extinction, we attempt to fit photometry and spectroscopy of 2MASS 1207 b with lower temperature thick-cloud model atmospheres from \citet{2011arXiv1102.5089M} (Section \ref{atmosphere}).  These models might be able to simultaneously explain 2MASS 1207 b's spectral shape and luminosity, in which case there would be no need to invoke either of the gray-extincting models.  This solution would also have the benefit of potentially explaining the whole class of under-luminous brown-dwarfs/giant exoplanets.

\section{Data and Observations\label{Observations}}

We summarize the relevant published photometry/spectroscopy of 2MASS 1207 in Table \ref{published}.  We also present new Gemini/T-ReCS mid-infrared data, described in Section \ref{Gemini/T-ReCS}, which allow us to put an upper-limit on the 8.7$\micron$ flux of 2MASS 1207 b.  We use these data to constrain disk models in Section \ref{Modeling}, shell models in Section \ref{Modeling DUSTY}, and brown-dwarf/giant-exoplanet model atmospheres in Section \ref{atmosphere}.

While photometry of 2MASS 1207 b has been published in a variety of near-infrared filters, multi-epoch data in any given filter rarely exists, the exception being NICMOS photometry, which was published for two epochs in the F090M and F160W filters \citep{2006ApJ...652..724S}.  However, other archival data do exist, that give more multi-epoch information than has previously been published.  For example, \citet{2005AA...438L..25C} verify the common proper motion of 2MASS 1207 A and b, but do not publish the photometry from this data.  We investigate 2MASS 1207 b's possible variability using archival VLT/NACO data in Section \ref{VLT/NACO}.  We also explore 2MASS 1207 A's variability using SMARTS/ANDICAM data that was previously published for parallax only \citep{2007ApJ...669L..41B}.  This is relevant to much of the previously published photometry of 2MASS 1207 b, which was calculated relative to 2MASS 1207 A and implicitly assumed 2MASS 1207 A was not variable.

\subsection{Gemini/T-ReCS 8.7$\micron$ Photometry\label{Gemini/T-ReCS}}
We observed 2MASS 1207 in queue-mode on 2008 March 29 UT and in classical-mode on 2010 March 31 UT and 2010 April 1 UT at Gemini-South, with its highly sensitive Thermal-Region Camera Spectrograph \citep[T-ReCS;][]{1998SPIE.3354..534T}.  We used the Si-2 filter ($\lambda_{central}=8.74\micron$), which is relatively insensitive to extinction/silicate absorption from a potential edge-on disk/shell, while being similarly sensitive as the N-band filter\footnote{\url{http://www.gemini.edu/sciops/?q=node/10085}}, and mostly immune to variations in precipitable water vapor \citep{2008SPIE.7016E..63M}.  Our data were taken in $\sim$320 s on-source blocks, which corresponds to $\sim$18 minutes clock-time when including chop-nod and other overheads.  These long integration sequences are required to build up enough S/N to shift and add on 2MASS 1207 A.  Combining the data in larger blocks appears to degrade our image quality due to guiding/nod-offset errors.

We reduced our data with the custom T-ReCS IDL software MEFTOOLS v. 5.0\footnote{\url{http://www.jim-debuizer.net/research/}}, which allows the user to interactively display individual T-ReCS saveset frames and remove those with bad electronic artifacts/noise properties.  We used this to discard approximately 2\% of our frames and produce combined $\sim$320 s chop-nod subtracted images.  

\begin{figure}
 \includegraphics[angle=0,width=\columnwidth]{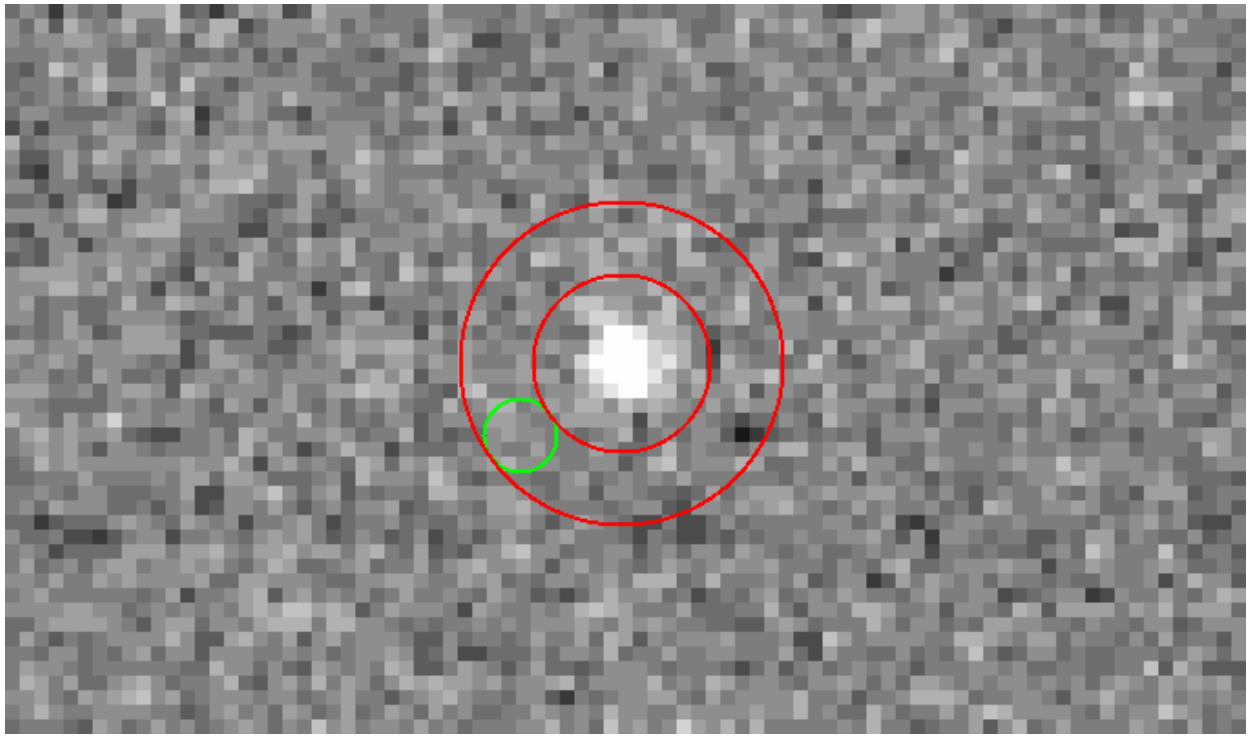}
\caption{Combined Gemini/T-ReCS Si-2 (8.7$\micron$) image of 2MASS 1207, where North is up and East is left.  2MASS 1207 A is at the center of the two red circles.  The near-infrared determined position of 2MASS 1207 b  \citep{2006ApJ...652..724S} is at the center of the green circle.  The halo subtracted region (described in Section \ref{Gemini/T-ReCS}) is the area between the 2 red circles ($r=$0.55" and $r=$1"), excluding the  2MASS 1207 b aperture, which is marked by the green circle ($r=$0.225").  We find with 50\% confidence (i.e. median) that 2MASS 1207 b's flux is $<$ 0.26 mJy, 84\% confidence (1$\sigma$) that 2MASS 1207 b's flux is $<$ 0.48 mJy and $\sim$99.9\% confidence (3$\sigma$) that 2MASS 1207 b's flux is $<$ 0.92 mJy.
\label{2MASS 1207_image}}
\end{figure}

For each of these images, we fit 2MASS 1207 A with a 2D Gaussian ellipsoid using the IDL software suite MPFIT \citep{2009ASPC..411..251M}.  We discard images where the fit centroid error on 2MASS 1207 A is $\ge$0.5 pixels (0.045").  This keeps our final combined image free of degradation from shift and add errors (T-ReCS' image quality is typically $\sim$0.3" FWHM).  Discarding frames with large centroid errors also has the benefit of selecting the frames with the best image quality (FWHM image quality cannot be accurately measured because of the low S/N on 2MASS 1207 A in a single frame).  Our selection leaves us with 15 $\sim$320 s blocks, which are then weighted by the S/N of the Gaussian ellipsoid fit, and combined.  While 14 out of 25 blocks from 2010 March 31 UT were usable, only one out of the 6 blocks from 2008 March 29 UT and none of the blocks from 2010 April 1 UT passed our centroid error cut.  The image quality from 2010 April 1 UT appears to have been significantly degraded by a nod-return/guiding error.  A summary of our observations and weather conditions is presented in Table \ref{observations}.  Our final combined image is 4749 s on-source (9498 s open-shutter, including chop-nod subtraction), and is shown in Figure \ref{2MASS 1207_image}.

Figure \ref{2MASS 1207_image} shows 2MASS 1207 A at the center of the two red circles, and a green circle at the near-IR determined position of 2MASS 1207 b \citep[sep=0.773", PA=125.37$^{\circ}$;][]{2006ApJ...652..724S}.  The measured FWHM of 2MASS 1207 A is 0.30", so the core of the 2MASS 1207 A point-spread-function (PSF) should have a negligible affect on the measured flux at the position of 2MASS 1207 b.  However, there does appear to be a slight increase in background at a separations $<$1", which is probably the result of 2MASS 1207 A's PSF seeing halo. We subtract off this small halo contribution by doing a median average of all pixels within a 2.5 pixel radius of the separation of 2MASS 1207 b, and at all position angles except for the aperture that we use for 2MASS 1207 b.

We used data from 2010 Mar 31 UT to perform an absolute flux calibration on 2MASS 1207 A using the mid-IR standard HR4888 \citep{1999AJ....117.1864C}, which is 16.011 Jy in the Si-2 filter\footnote{\url{http://www.gemini.edu/sciops/instruments/midir-resources/imaging-calibrations/michelle-std-fluxes}, which is calculated for Gemini-N/Michelle, but the Si-2 filters of Michelle and T-ReCs are similar}.  The best 4-hour period of 2MASS 1207 data (which included 9 of our 15 usable frames) was averaged, and compared to images of HR4888 taken immediately after (HR4888 was found to vary by $\sim$5\% over the 3 calibration images).  We use a 15 pixel (1.35") aperture and find that 2MASS 1207 is 0.000343 times as bright as HR4888 in the Si-2 filter, which corresponds to an absolute flux of 5.49 mJy.  We assign an 8 \% error to this measurement (0.44 mJy), which is conservative given the fidelity of the Si-2 filter and the night's good photometric quality.  This value is consistent with published values of 5.6 mJy $\pm$1 mJy in the Si-2 filter \citep{2004AA...427..245S}, 5.74 mJy in the slightly bluer Spitzer 8$\micron$ filter \citep{2006ApJ...639L..79R}, Spitzer spectroscopy \citep{2008ApJ...676L.143M}, and roughly consistent with our own measurements using the 2008 Mar 29 UT data (4.3 mJy $\pm$0.5 mJy), which suggests that 2MASS 1207 A is probably not wildly variable at these wavelengths, despite the presence of a near edge-on disk.
 
Using 2MASS 1207 A as a PSF, we do a best-fit to determine how much flux is at the near-IR determined position of 2MASS 1207 b \citep[sep=0.773", PA=125.37$^{\circ}$;][]{2006ApJ...652..724S}.  We assume Gaussian error bars (equal on all pixels for the background limited case), and fit within a 5 pixel diameter aperture (see Figure \ref{2MASS 1207_image}) to avoid contamination by 2MASS 1207 A.  We do a Monte Carlo evaluation of the background fluctuations by doing a similar aperture measurements at 10000 randomly chosen position angles and separations $>$ 1.5" from 2MASS 1207 A to avoid residual halo contributions.  For each Monte Carlo trial, we subtract the flux of a randomly chosen background aperture from the measured flux at the position of 2MASS 1207 b.  This gives us an estimate, for each trial, of what the true flux of 2MASS 1207 b might be.  We discard all negative flux measurements based on the Bayesian prior that 2MASS 1207 b's flux must be positive, and use the remaining cases to construct a cumulative distribution function, which calculates the probability that 2MASS 1207 b's flux is less than a given value (but greater than 0).  We include an 8\% absolute calibration error (see the previous paragraph), although increasing this error up to $\sim$20\% has a negligible effect on our results.  We find with 50\% confidence (i.e. median) that 2MASS 1207 b's flux is $<$ 0.26 mJy, 84\% confidence (1$\sigma$) that 2MASS 1207 b's flux is $<$ 0.48 mJy and $\sim$99.9\% confidence (3$\sigma$) that 2MASS 1207 b's flux is $<$ 0.92 mJy.

\subsection{VLT/NACO Near-Infrared Photometry\label{VLT/NACO}}
2MASS 1207 A and b have been spatially resolved by VLT/NACO at several epochs.  Some of these data have been published as astrometry only, and some have not been published at all.  Here we use archival data to investigate the possibility that 2MASS 1207 b is photometrically variable.  

2MASS 1207 was observed 6 times in the Ks filter between 2004 April 27 UT and 2009 June 30 UT.  Shorter wavelength filters were also used at multiple epochs, but generally on the same nights as Ks, so variability information from these filters is redundant.  Shorter wavelength filters also suffer from lower signal-to-noise and a worse adaptive optics correction. Therefore we only focus on the Ks-band data here.

Of the 6 epochs of Ks data, 2 dates (2005 Mar. 30 and 2005 May 18) suffered from poor seeing so that we could not produce reliable resolved photometry of 2MASS 1207.  The other four epochs were reduced using the custom IRAF pipeline described in \citet{2003ApJ...587..407C}.  We performed PSF-fitting photometry on 2MASS 1207 A and b, and a background star (2MASS 12073400-3932586), using IRAF/\textit{daophot} \citep{1987PASP...99..191S} and \textit{allstar}.  Seeing conditions on 2009 June 30 were marginal, so the data were reduced by aperture photometry and assigned conservative errors.  A summary of our observations and photometry is presented in Table \ref{NACO observations}.  Ignoring the low-quality 2009 June 30 data, we find that compared to the background star, 2MASS 1207 A is variable by, at most, $\sigma=$0.05 mag (0.09$\pm$0.03 mag peak-to-peak), which is similar to our optical SMARTS/ANDICAM results in Section \ref{SMARTS/ANDICAM}.  2MASS 1207 b is variable by, at most, $\sigma=$0.12 mag (0.23$\pm$0.07 mag peak-to-peak), which is not abnormally high for a young object with star-spot and/or accretion variability.

\subsection{SMARTS/ANDICAM I-band Photometry\label{SMARTS/ANDICAM}}
The combined system, 2MASS 1207 A and b was observed by \citet{2007ApJ...669L..41B} between January 2006 and April 2007 with SMARTS/ANDICAM to measure the parallax distance to 2MASS 1207.  Here we present the photometric results of this data set, which provide 53 nights of I-band photometry, taken monthly, in groups of several nights.  Observational details are described in \citet{2007ApJ...669L..41B}.  Special care was taken to always position the target on the same CCD pixel, and all data were taken near transit to aid the interpretation of the parallax data.  Relative photometry between 2MASS 1207 and the 12 brightest stars in the field gives an average variability of $\sigma=0.03$ mag.  Given that the differential magnitude between 2MASS 1207 A and b is 7.8 in the similar NICMOS F090M filter \citep{2006ApJ...652..724S}, this variability must be completely dominated by 2MASS 1207 A's intrinsic variability.  The variability is smaller than the error bars of published relative photometry of 2MASS 1207 b \citep[see for example][]{2004AA...425L..29C}, which legitimizes 2MASS 1207 A's use as a photometric calibrator for 2MASS 1207 b (although as we describe in Section \ref{VLT/NACO}, the nearby background star 2MASS 12073400-3932586 is a better calibrator for variability studies).

\begin{figure}
 \includegraphics[angle=90,width=\columnwidth]{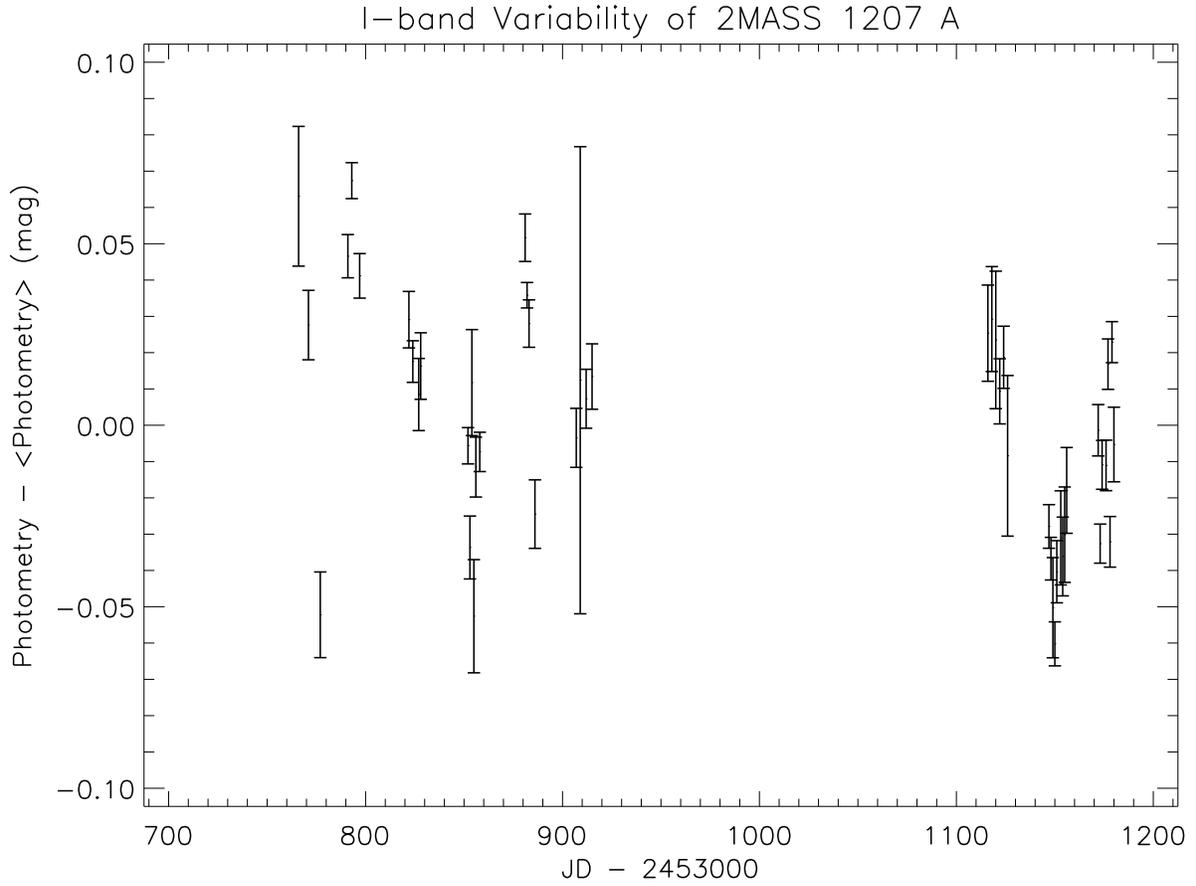}
\caption{SMARTS/ANDICAM I-band variability of 2MASS 1207 A.  2MASS 1207 A is variable with a modest $\sigma\simeq$3\%.  This means 2MASS 1207 A is a legitimate photometric calibrator for 2MASS 1207 b.\label{A variability}}
\end{figure}

Variations in the I-band magnitude of 2MASS 1207 with respect to its average I-band value are plotted in Figure \ref{A variability}.  Error bars are calculated empirically from the multiple frames taken on each individual night.

\section{The Near Edge-on Disk Hypothesis\label{Modeling}}
A possible explanation for 2MASS 1207 b's under-luminosity is that it is being partially extincted by a near edge-on disk of large (gray-extincting) dust grains \citep{2007ApJ...657.1064M,2010AA...517A..76P}.  Such a disk must produce $\sim$2.5 mags of extinction at J-band, with nearly gray extinction.  The disk must create only moderate photometric variability.  The disk may not have an extremely unlikely viewing angle/geometry.  And the disk properties should be consistent with our current picture of disk evolution.  Additionally, any dust around 2MASS 1207 b will have its emission constrained by our 8.7 $\micron$ photometry.  In this section, we discuss the viability of the near edge-on disk hypothesis, with respect to these conditions.

\subsection{RADMC Disk Models\label{RADMC Disk Models}}
We used the RADMC radiative transfer code \citep[]{2004AA...417..159D} and RAYTRACE, a post-processing tool, to calculate the emerging spectral energy distribution (SED) of a hypothetical disk around 2MASS 1207 b.  The primary input parameters of the RADMC/RAYTRACE codes are the SED and mass ($M_{\rm star}$) of a central source, and the inner radius ($R_{\rm in}$), outer radius ($R_{\rm disk}$), inclination ($i$), and mass ($M_{\rm disk}$) of its disk.  Other relevant parameters include the disk's vertical flaring geometry (described below), and a dust grain distribution with associated optical properties.

As we will show later in this section, our Gemini/T-ReCS 8.7$\micron$ photometry upper-limit is not sensitive enough to detect the presence of a disk around 2MASS 1207 b.  Consequently, there is no data available at wavelengths longer than L-band to distinguish between different possible disk models.  However, 2MASS 1207 A has a considerable amount of data at longer wavelengths.  As such, our strategy is to model the disk around 2MASS 1207 A first, and scale the model for use around 2MASS 1207 b.

For 2MASS 1207 A, we adopted a $T_{\rm eff}$=2600 K, $R_{\rm A}$=0.25 $R_{\rm \odot}$, $log(g)$=4 and $M_{\rm A}$=0.025 $M_{\rm \odot}$ AMES-DUSTY synthetic spectrum \citep{2001ApJ...556..357A} for the central source, similar to the best-fit model determined by \citet{2007ApJ...657.1064M}.  For the disk parameters, we assume $R_{\rm in}$=3.3 $R_{\rm A}$ \citep[the dust sublimation radius for 2MASS 1207 A;][]{2008ApJ...676L.143M}, $R_{\rm disk}$=20 AU \citep[$\sim$0.46$\times$projected binary separation;][]{1994ApJ...421..651A}, and $M_{\rm disk}$=0.01 $M_{\rm A}$.  We note the precise values of $R_{\rm disk}$ and $M_{\rm disk}$ have a negligible effect on the SED for $\lambda<$20$\micron$.  

We vary the model's dust grain distribution, vertical geometry, and disk inclination to determine a best-fit to the data described in Section \ref{Observations}.  For the dust grain distribution we applied a single olivine dust species \citep{2003AA...408..193J} and varied the minimum diameter ($a_{\rm min}$) of the grains between 0.1$\micron$ and 10$\micron$.  We assumed a dust grain size power-law with $n(a)\propto a^{-3.5}$ \citep{1977ApJ...217..425M}, and a maximum dust grain size of 1 mm.  For the vertical geometry, we ran two sets of models: a fully-flared model (hereafter, ``flared"), where RADMC calculates the disk-height self-consistenty (assuming hydrostatic equilibrium and dust-gas coupling), and a set of flatter (hereafter, ``flat") models where the disk scale heights are parameterized.  The flat disk models are assumed to follow a power-law description with $\frac{H_{p}}{r}\propto r^{1/7}$ and outer disk pressure scale heights ($H_{p}/r$) ranging from 0.04 to 0.10 at $R_{\rm disk}$.  The final free parameter, inclination, is varied between 5$^{\circ}$ and 90$^{\circ}$ to determine a best-fit.

\begin{figure}
 \includegraphics[angle=0,width=\columnwidth]{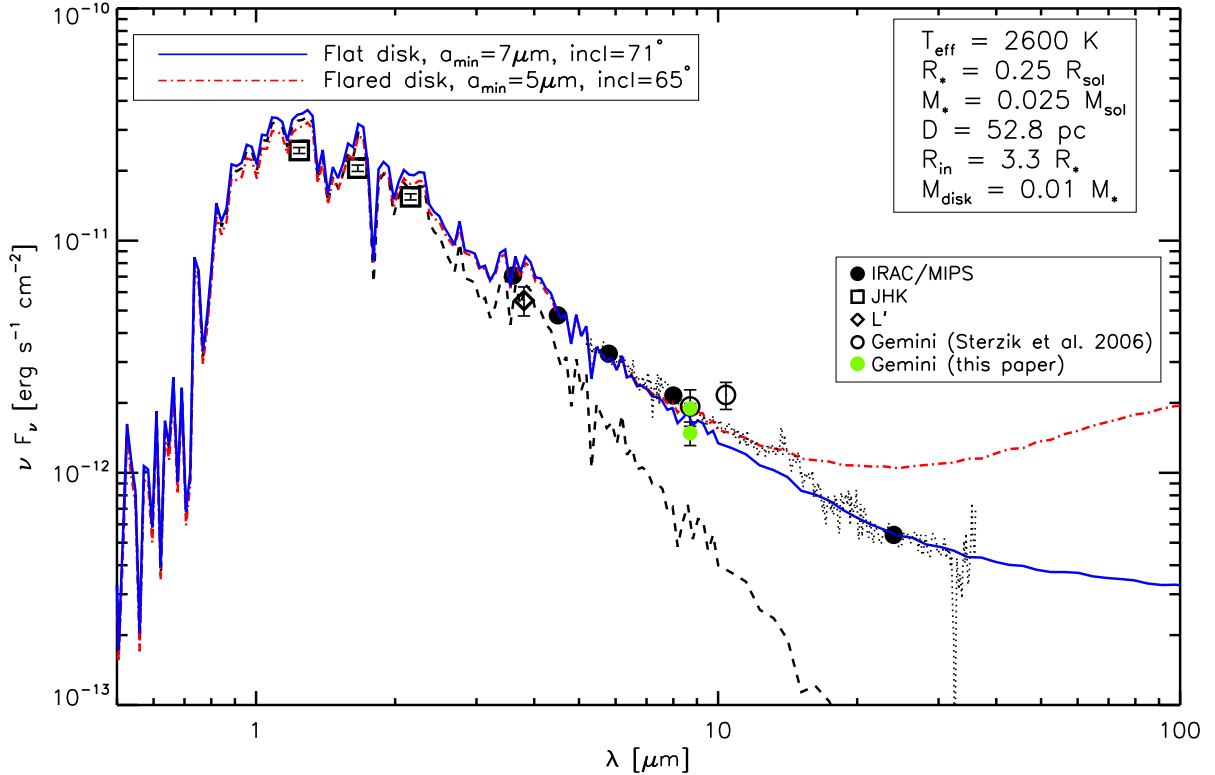}
\caption{Best-fit flared and flat disk models for 2MASS 1207 A.  The black, dashed curve is the AMES-DUSTY model spectrum for 2MASS 1207 A's photosphere, the red, dot-dashed curve  is the flared disk model, the blue, solid curve is the flat disk model, and the black dotted curve is the Spitzer IRS spectrum \citep{2008ApJ...676L.143M}.  Both models adequately fit the near-infrared data ($\lesssim10\micron$), but only the flat model adequately fits the mid-infrared data ($\gtrsim10\micron$).  Thus, we conclude that the flat disk model is better for 2MASS 1207 A, and along with evidence that flat disk models are better for most brown dwarfs \citep{2010ApJ...720.1668S}, we use this to justify modeling 2MASS 1207 b's hypothetical disk as flat.
\label{A RADMC}}
\end{figure}

The flared disk models do not fit the observed SED of 2MASS 1207 A.  While the models can fit the near-IR observations (assuming inclinations $<$65$^{\circ}$), they predict significantly higher fluxes  than are observed in the mid-infrared (See Figure \ref{A RADMC}).  The best-fit flat disk model ($H_p/r$=0.06, $a_{min}$=7$\micron$ and $i$=71$^{\circ}$), also shown in Figure \ref{A RADMC}, provides an adequate fit to all of the 2MASS 1207 A data, although other parameter combinations also result in acceptable fits.    Exploring the degeneracies in these parameters, we find that the disk around 2MASS 1207 A has 0.06$<H_{\rm p}/r<$0.08, $a_{min}>$5$\micron$, and 70$^{\circ}<i<75^{\circ}$.  The best-fit flat models are consistent with the results of \citet{2008ApJ...676L.143M}. Both models suggest advanced dust processing and dust settling in the disk around 2MASS 1207 A.

\begin{figure}
 \includegraphics[angle=0,width=\columnwidth]{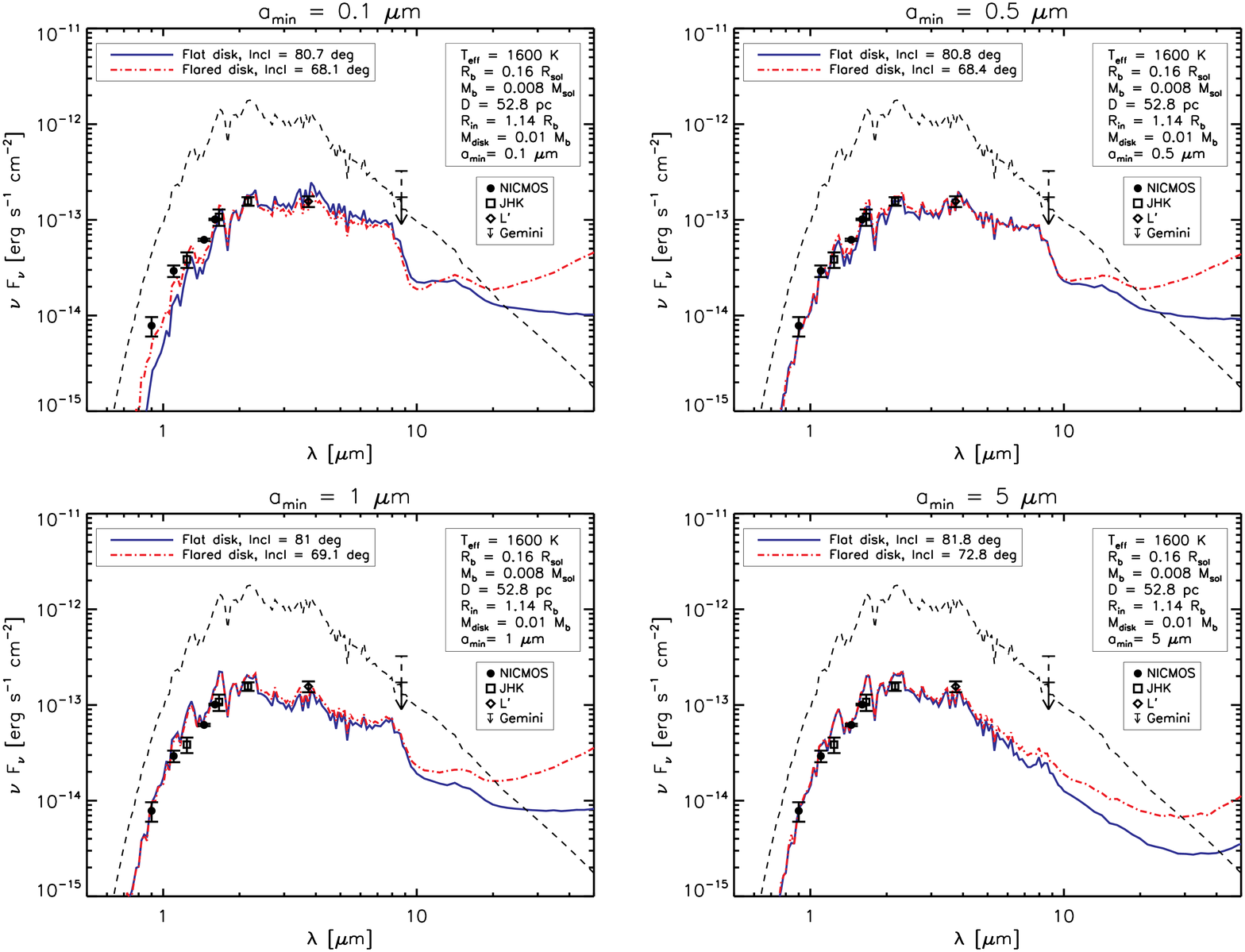}
\caption{Best-fit flared and flat disk models for 2MASS 1207 b using four minimum grain sizes.  The black, dashed curves are the AMES-DUSTY model for 2MASS 1207 b's photosphere (no extinction), the red, dot-dashed curves are the flared disk models, and the blue, solid curve are the flat disk models.  Note that our 8.7$\micron$  upper-limit is drawn with a downward arrow at the median value connected by a solid line to a horizontal +1$\sigma$ upper-limit, which is connected by a dashed line to a horizontal +3$\sigma$ upper-limit.  The flared and flat models are indistinguishable at near-infrared wavelengths ($\lesssim10\micron$), and no data exists at longer wavelengths.  The smallest minimum grain size (0.1$\micron$) flat-disk models cannot simultaneously fit the bluest and reddest data.  For grain sizes $\geq$0.5$\micron$, the models fit the data adequately.
\label{B RADMC}}
\end{figure}

\begin{figure}
 \includegraphics[angle=0,width=\columnwidth]{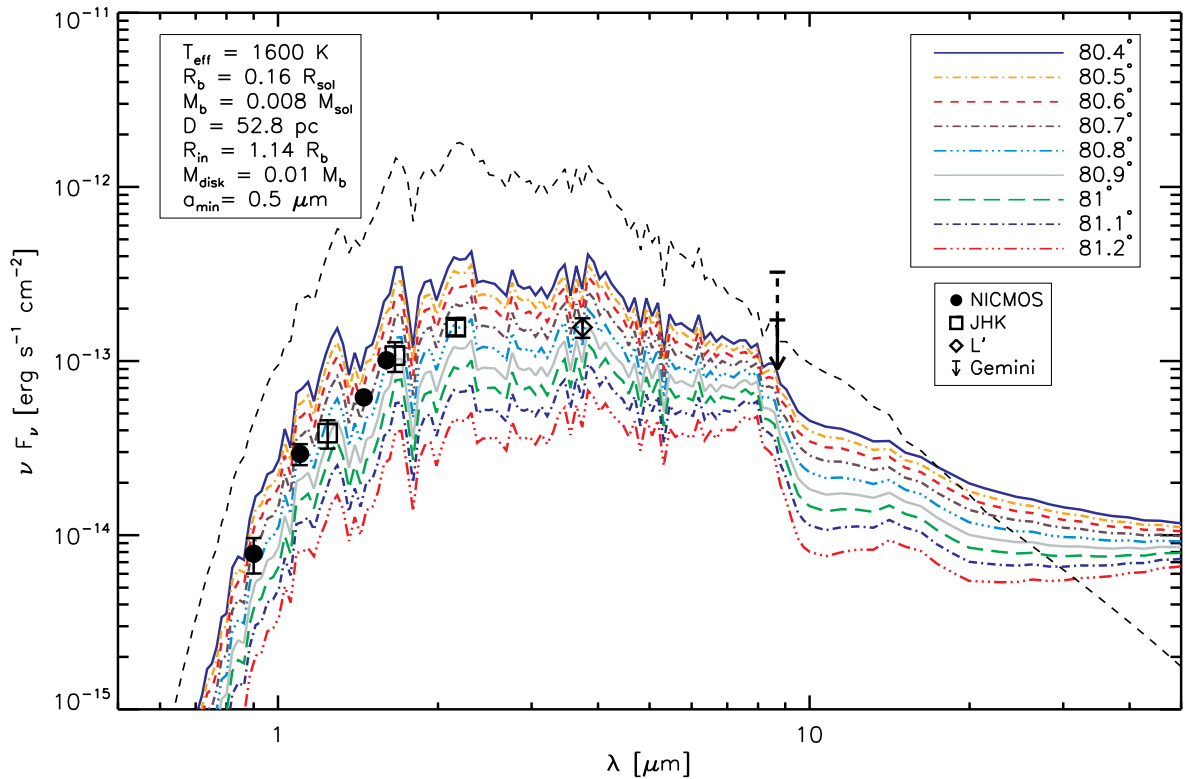}
\caption{$a_{\rm min}=0.5\micron$ flat disk models for 2MASS 1207 b.  For a given model, inclination is very tightly constrained to about $\sim$0.3$^{\circ}$ to fit the observed SED.  Small non-axisymmetric structures (such as warps, gaps, hydrodynamic fluctutations and clumpiness) in a hypothetical disk around 2MASS 1207 b, which are analogous to small inclination changes, would likely cause variability larger than has been observed (See Section \ref{variability}).
\label{inclination}}
\end{figure}

To model the hypothetical disk around 2MASS 1207 b, we scaled down the flared and flat disk models of 2MASS 1207 A.  For the central source we use an AMES-DUSTY model with $T_{\rm eff}$=1,600 K, $R_{\rm A}$=0.16 $R_{\rm \odot}$ and $log(g)$=4.5 \citep{2010AA...517A..76P}.  We used the relevant disk parameters from 2MASS 1207 A: $H_{\rm p}/r$=0.06, $M_{\rm disk}$=0.01 $M_{\rm b}$, $R_{\rm in}$=1.14 $R_{\rm b}$ (the dust sublimation radius) and $R_{\rm disk}$=10 AU \citep[$\sim$0.2$\times$projected binary separation;][]{1994ApJ...421..651A}.  Our best-fit flared and flat models are shown for 4 different values of $a_{\rm min}$ in Figure \ref{B RADMC}.  The lack of ($\lambda>$10$\micron$) mid-infrared data makes it impossible to distinguish between the flared and flat disk models, although we note that disks around low-mass stars/brown dwarfs are found to be flatter than disks around higher-mass stars \citep{2010ApJ...720.1668S}.  We varied $a_{\rm min}$ from 0.1$\micron$ to 5$\micron$ and find that the $a_{\rm min}$=0.1$\micron$ flat-disk models cannot  simultaneously fit the $K_{\rm s}$ and $L'$, and NICMOS observations.  Models with $a_{\rm min}\ge$0.5$\micron$ fit the data adequately, although the disk models with $a_{\rm min}=1\micron$ or 5$\micron$ do not fit the data quite as well as the $a_{\rm min}=0.5\micron$ due to the redness observed between L' and the shorter wavelength data. In each case, the models have been fit by varying inclination, which is constrained by our requirement that the disk extinct the central source by $\sim$2.5 mags at J-band in order to fit 2MASS 1207 b's near-infrared photometry.  Figure \ref{inclination} shows the $a_{min}=$0.5$\micron$ flat model from Figure \ref{B RADMC}, but with varying inclination.  This model is consistent with the measured data for an inclination range of 80.6$^{\circ}<i<80.9^{\circ}$.  Different disk models have similarly tightly constrained inclinations, although the actual ranges vary between models.  For a given model, the inclination of the disk (or non-axisymmetric structure) cannot significantly vary in time without producing strong variability (see Section \ref{variability}).

For the remainder of this paper we use the $a_{min}=$0.5$\micron$ models for 2MASS 1207 b.  We continue to discuss the flared and flat models, although we prefer the flat models due to the presence of one around 2MASS 1207 A, and more generally, because of the ubiquity of flat disks around low-mass stars/brown-dwarfs \citep{2010ApJ...720.1668S}.

\subsection{Variability in Near Edge-on Disks\label{variability}}
The lack of near-infrared variability exhibited by 2MASS 1207 b would be unusual for a system with a partially extincting, near edge-on disk. Young systems with edge-on or near edge-on disks have been shown to exhibit variability over a variety of masses (brown-dwarfs, T Tauri stars, Herbig Ae-Be stars).  This variability is generally the result of eclipsing non-axisymmetric structures in Keplerian orbit, such as warps, hydrostatic fluctuations, gaps, spirals, and clumpiness.  The best studied edge-on variables are UX Ori type stars, which display frequent eclipse-like extinction events caused by hydrodynamic fluctuations along their puffed-up inner-rims.  However, the UX Ori phenomenon is thought to only exist in completely self-shadowed disks, which are usually around Herbig Ae-Be stars \citep{2003ApJ...594L..47D}.  In lower-mass T-Tauri stars and brown dwarfs, disk flaring causes the outer-parts of the disk to be the dominant source of extinction \citep{2003ApJ...594L..47D}.  The structural sources of variability in the outer parts of disks are less well understood than inner-rim variability.  A far-away companion, mis-aligned with the plane of the disk (i.e. 2MASS 1207 A) can cause warps in the outer-parts of the disk \citep{2010AA...511A..77F}, but in this case the warp will orbit with the period of the binary, which is much longer than the orbital period of the disk, and is unlikely to cause short-term variability.  Other sources of non-axisymmetric structure have been studied with regard to specific perturbations \citep{2010ApJ...719.1733F,2011arXiv1103.0781F}, but a general theory does not yet exist.

Empirically, we know that edge-on and near edge-on disks that extinct their host star exhibit variability over many timescales.  Famous examples include the T Tauri stars AA Tau, which exhibits $\sim$1 mag variability on periods of days/weeks \citep{1999AA...349..619B,2003AA...409..169B}, HH 30, which shows dramatic variability in spatially resolved scattered light images \citep{1999ApJ...516L..95S,2007AJ....133..845W}, T Tau Sa, which varies in brightness by several magnitudes on timescales of years \citep{2008ApJ...676.1082S,2010AA...517A..16V}, and UY Aur B, which became fainter by several magnitudes over a $\sim$50 year period \citep{1944PASP...56..123J,1995ApJ...444L..93H}.  A systematic analysis of NGC 2264 by \citet{2010AA...519A..88A} revealed that 28\%$\pm$6\% of classical T Tauri stars exhibit AA Tau-like variability with amplitudes as large as 137\% over a 23 day period.  \citet{2010AA...519A..88A} do not publish the variability amplitude of all of their AA Tau-like systems, so we do not know the lower-limit amplitude.  However, their result systematically confirms that AA Tau-like variability is common, and probably synonymous with near edge-on/edge-on disks.

The physical mechanisms leading to disk asymmetries vary as a function of spectral type and accretion rate and thus the well-studied disks described in the previous paragraph may not be analogous to brown dwarfs.  Although examples are sparse, several groups \citep{2007ApJ...666.1219L,2009MNRAS.398..873S,2010ApJ...714...45L,2010AJ....140.1486L} report evidence for variable occultation of brown dwarfs from edge-on or near edge-on disks.  These cases provide empirical examples, but unbiased, systematic variability studies of near edge-on/edge-on disks around brown dwarfs are rare.

From a modeling perspective, a small warp of just 1 degree (which we approximate as changing the inclination of our RADMC disk by 1 degree) would cause variability of amplitude 2.20 mags in the NICMOS F160W filter (as well as 2.13 in the NICMOS F090W filter, and 2.16 in Ks-band).  This is inconsistent with measured variability amplitudes of 0.03$\pm$0.03 in F160W, 0.24$\pm$0.5 mag in F090W, and 0.23$\pm$0.07 in Ks-band as well as the general stability of the multi-epoch, multi-wavelength photometry and spectroscopy summarized in Table \ref{published}.  With respect to the variable brown dwarfs discussed above, which we caution is not an unbiased sample, \citet{2010ApJ...714...45L} and \citet{2010AJ....140.1486L} find an amplitude of variability of 1.3 mags and 0.9 mags at Ks for TWA 30A and TWA 30B over a period of months, \citet{2009MNRAS.398..873S} find an amplitude of variability of 0.45 mags at K for their ``Object \#2" over a period of about a week and \citet{2007ApJ...666.1219L} find an amplitude of variability of $\sim$0.5 mags at Ks for 2MASS 0438+2611 over 2 epochs separated by 6 years.  All of these objects are substantially more variable than 2MASS 1207 b.

As a result, we find the existence of a near edge-on disk around 2MASS 1207 b to be incompatible with 2MASS 1207 b's measured photometric stability.  However, we caution that this conclusion is based on an incomplete knowledge of the structure and variability of extincting disks around brown dwarfs.  Further studies that catalog lightcurves of many young low-mass stars (such as the ongoing Spitzer/YSOVAR program, and future surveys, such as PAN-STARRS and LSST) will better constrain the expected variability of extincting disks around young brown dwarfs.

\subsection{The Expected Frequency of Partially Extincting Disks Around Young Brown Dwarfs}
\textit{A priori}, it would be unlikely (and perhaps unfortunate) that 2MASS 1207 b, a system which is currently unique as the closest, youngest, extremely low-mass object, is inclined to our line of sight in a near edge-on configuration where it is ``partially" extincted.  In this section, we investigate the probability that a brown dwarf system is in a configuration such that its disk ``partially" extincts the central source.  We use our RADMC disk models for 2MASS 1207 b and calculate the extinction produced by the hypothetical disk when viewed at different inclinations varying from 0 to 90 degrees.  

\begin{figure}
\begin{center}
 \includegraphics[angle=0,width=0.7\columnwidth]{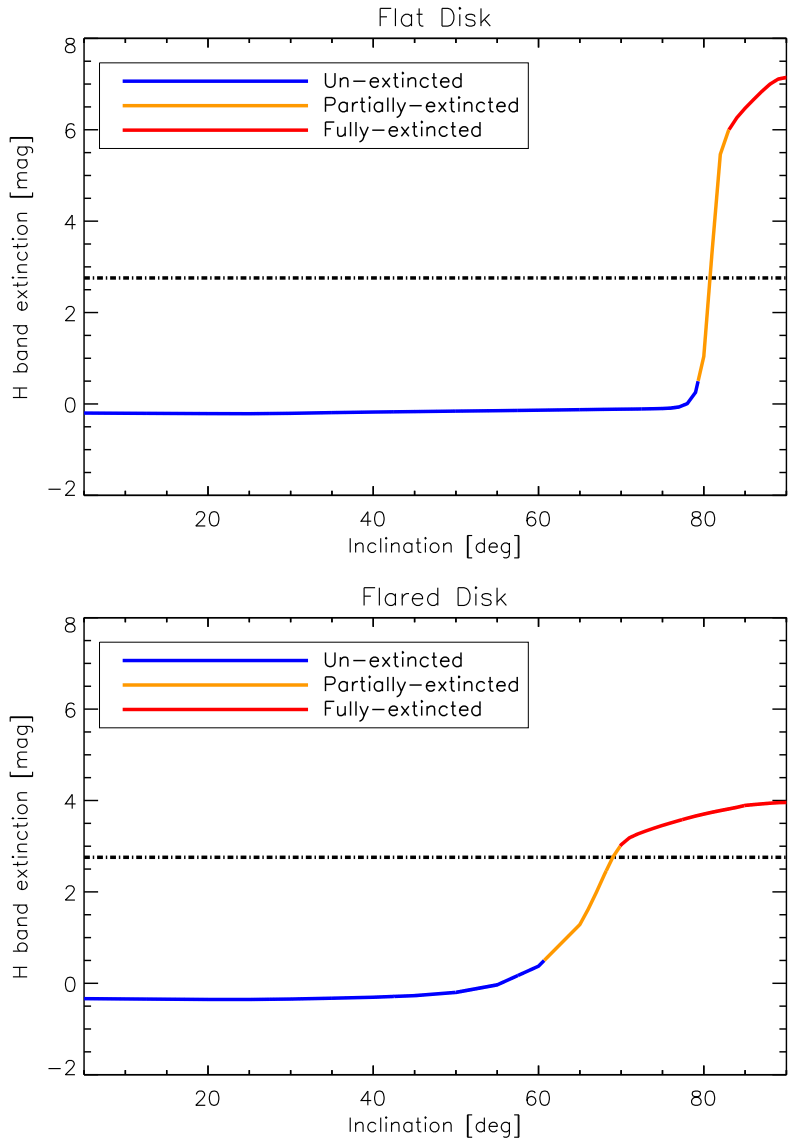}
\end{center}
\caption{The under-luminosity of a brown dwarf extincted by its disk as a function of luminosity, shown for our flat disk model, and our flared disk model (the under-luminosity of 2MASS 1207 b is marked as a horizontal, dot-dashed line).  For low inclinations, the disk does not extinct the star.  Eventually, the outer edge of the disk moves into the line of sight, and extinction rises rapidly.  Finally, at very high inclinations, the under-luminosity is capped by the scattered light emission of the disk.  We term these regions \textit{un-extincted}, \textit{partially-extincted}, and \textit{fully-extincted}.  That 2MASS 1207 b is in the partially-extincted region is, \textit{a priori}, unlikely, but not statistically refutable.  If a disproportionate number of objects in 2MASS 1207 b's class are found to be partially-extincted, then there must be another source of their under-luminosity.
\label{extinction}}
\end{figure}

In Figure \ref{extinction} we show the H-band extinction predicted by the models as a function of disk inclination, for both flat and flared geometries.  We note that there are 3 distinct regions of each plot: an \textit{unextincted} region, a \textit{partially extincted} region where the extinction rises sharply, and a \textit{fully extincted} region where the extinction levels off.  Physically, the partially extincted region begins when the outer edge of the disk moves into our line of sight towards the brown dwarf.  As inclination continues to increase, the extinction towards the brown dwarf rises until the fully extincted region, where scattered light (which is assumed to be isotropic in RADMC) caps extinction even as opacity continues to rise.  

Based on this analysis, and assuming an isotropic distribution of disk inclinations, there is a 81.5\% probability that a brown dwarf with our flat disk model is unextincted ($A_{H}=$0-0.5 mags), a 6.4\% probability that it is partially extincted ($A_{H}=$0.5-6.0 mags), and a 12.1\% probability that it is fully extincted ($A_{H}=$6.0-7.1 mags).  For the flared disk model, there is a 51.0\% probability that it is unextincted ($A_{H}=$0-0.5 mags), a 14.6\% probability that it is partially extincted ($A_{H}=$0.5-3 mags), and a 34.4\% probability that it is fully extincted ($A_{H}=$3-4 mags).  For brown dwarfs, the flat disk model is more accurate than the flared disk model (see Section \ref{RADMC Disk Models}). Although the exact analysis depends strongly on disk flaring, the two cases presented here bracket the likely geometries.

Assuming 2MASS 1207 b has a disk, but ignoring its under-luminosity as prior evidence for why it should have a near edge-on disk, the probability that 2MASS 1207 b's disk is in the inclination range that causes partial extinction is $\sim$6.4\%.  This also neglects the fact that 2MASS 1207 A's disk is thought to be near edge-on, and disks in binaries tend to be aligned.  \citep{2004ApJ...600..789J,2006AA...446..201M}.  Because 2MASS 1207 b is such a unique object, and because we do not yet understand the likelihood that alternative effects are causing its under-luminosity, we cannot use the fact that 2MASS 1207 b's hypothetical disk is at an unlikely inclination to reject the disk's existence\footnote{Following Bayes' theorem, $P(edge-on|under-luminous)=\frac{P(under-luminous|edge-on)P(edge-on)}{P(under-luminous).}$.  To calculate the probability that a brown dwarf has a near edge-on disk, given that it is under-luminous, we need to know the probability that brown dwarfs are under-luminous (the denominator term).  Because we do not understand the likelihood that alternative effects are causing the under-luminosity, and because we do not know what percentage of brown dwarfs are under-luminous, we do not know $P(under-luminous)$, so we cannot evaluate $P(edge-on|under-luminous)$}.  However, when more objects like 2MASS 1207 b (i.e. young, low-mass brown dwarfs/exoplanets) are found by surveys like WISE, our models predict that 81.5\% will be unextincted, $\sim$6.4\% will be under-luminous by 0.5-6.0 mags, and 12.1\% will be under-luminous by 6.0-7.1 mags.  A higher percentage in the middle category would imply that disk inclination cannot explain the under-luminosity phenomenon.

\subsection{Other Under-luminous Brown-Dwarfs/Exoplanets}
While the previous section predicts the percentage of objects in 2MASS 1207 b's class that should appear under-luminous (partially extincted) due to the presence of a near edge-on disk, there are no other objects quite like 2MASS 1207 b in terms of its youth and low mass.  However, there are other known under-luminous brown dwarfs and giant planets (e.g. HD 203030 B, HN Peg B and HR 8799 bcde), which are known to be older than 2MASS 1207 b.  Since the gas-rich, geometrically thick disks around these older systems will have dissipated \citep[and in the case of HR 8799, the disks would more likely be face-on;][]{2008Sci...322.1348M}, the near edge-on disk hypothesis proposed for 2MASS 1207 b would not be applicable.  In the previous sections, we have stated why we find the 2MASS 1207 b near edge-on disk hypothesis to be unlikely.  A further reason is that there are other under-luminous brown dwarfs that cannot be explained by a near edge-on disk, and thus, there is no reason to invoke one to explain the behavior of 2MASS 1207 b.

Are there other patterns that might be shared between these under-luminous brown dwarfs?  \citet{2009ApJ...705L.204M} note that all of the known under-luminous objects are young L/T transition brown-dwarfs (although \citet{2010AA...517A..76P} describe 2MASS 1207 b as an early L-type), and this pattern (sometimes parsed as the gravity dependence of the L/T transition) has been studied by numerous groups, including \citet{2008ApJ...689..436L}, \citet{2009ApJ...699..168D}, \citet{2009ApJ...702..154S}, \citet{2008ApJ...689.1327S} and \citet{2010ApJ...723..850B}.  As brown dwarfs cool from L-type to T-type, their dusty clouds are thought to dissipate \citep{2002ApJ...571L.151B} and settle below the brown dwarf photosphere.  This pattern implies that non-equilibrium chemistry and dust-cloud physics might play a crucial role in explaining these systems' apparent under-luminosities.

A possible explanation for the under-luminosity of the HR 8799 planets has recently been proposed by \citet{2011arXiv1101.1973C} and \citet{2011arXiv1102.5089M}, who demonstrate that a thick cloud atmosphere can reproduce the observed photometry of the HR 8799 planets.  The thick cloud models might be applicable to 2MASS 1207 b as well, and we investigate this possibility in Section \ref{atmosphere}.

\section{The Dust Shell Hypothesis}\label{Modeling DUSTY}
The problems in invoking a near edge-on disk to explain the under-luminosity of 2MASS 1207 b stem from the fact that near edge-on and edge-on disks tend to produce more variability than is detected for 2MASS 1207 b, along with the fact that such a configuration is \textit{a priori} unlikely and cannot be used to explain other under-luminous brown dwarfs.  In this section, we model the system with a shell of dust, which should provide the extinction of a near edge-on disk, without the geometric effects that we have previously deemed unlikely.  Such a shell may have been discovered around the brown dwarf, G 196-3 B \citep{2010ApJ...715.1408Z}.

We use the dust shell radiative transfer code, DUSTY\footnote{User Manual for DUSTY, University of Kentucky Internal Report, accessible at \url{http://www.pa.uky.edu/~moshe/dusty}}$^{,}$\footnote{Regrettably, the DUSTY dust-shell radiative transfer code shares a name with the AMES-DUSTY brown-dwarf model atmospheres we use in this paper.  So to be clear, in this section, we are enshrouding an AMES-DUSTY brown-dwarf model atmosphere with a DUSTY dust-shell.  Throughout this paper, we refer to the brown-dwarf models atmosphere's ``AMES-DUSTY" (although they are often referred to by other authors as DUSTY), and the dust-shell modeling software as ``DUSTY".} \citep{1999astro.ph.10475I}, to model dust shells at a variety of distances from the central brown dwarf.  In all cases, the central source is the AMES-DUSTY\footnotemark[\value{footnote}] brown dwarf model atmosphere ($T_{\rm eff}$=1,600 K, $R_{\rm b}$=0.16 $R_{\rm \odot}$ and $log(g)$=4.5) used as a model of 2MASS 1207 b in Section \ref{RADMC Disk Models}.  We also use the same dust grain size distribution and optical properties as described in \ref{RADMC Disk Models}, with a minimum grain-size of $a_{min}$=1$\micron$ (although varying this does not significantly change our results).  For each shell, we fix the J-band extinction to be 2.5 magnitudes (the observed under-luminosity of 2MASS 1207 b).  The shells are geometrically thin, and placed at $R_{\rm b}$, 5$R_{\rm b}$, 10$R_{\rm b}$, 20$R_{\rm b}$, 40$R_{\rm b}$, and 80$R_{\rm b}$.  Since DUSTY is a 1-D code, we are forced to choose a slab geometry (which we use for the $R_{\rm b}$ model), and a shell geometry (which we use for all others), where the central source is assumed to be a point-source.  The difference between using a slab and point-like approximation is small compared to the bulk SED shape, so we consider the point-like approximation reasonable for our purposes.  Another limitation of this modeling approach is that a shell of warm dust right near the surface of the brown dwarf would heat the brown dwarf and change the output spectrum.  Thus our closest-in shells, while useful for modeling the mid-infrared output, are somewhat non-physical, especially in the near-infared, where self-consistent models would be necessary to produce the correct spectral features.

\begin{figure}
 \includegraphics[angle=0,width=\columnwidth]{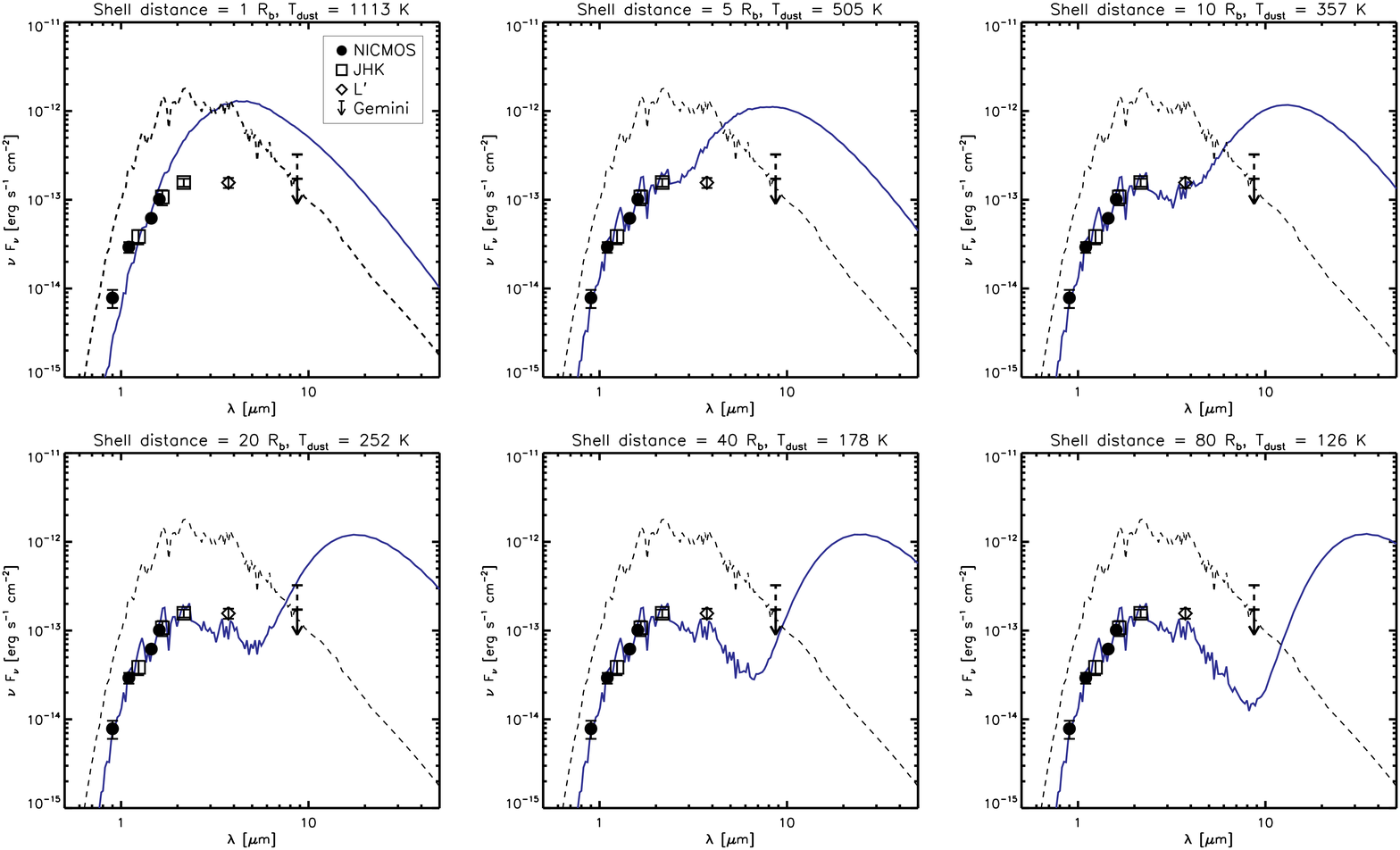}
\caption{Models of 2MASS 1207 b processed through a spherical shell of dust at different radii (1$R_{\rm b}$, 5$R_{\rm b}$, 10$R_{\rm b}$, 20$R_{\rm b}$, 40$R_{\rm b}$ and 80$R_{\rm b}$).  The dashed, black curves are the unattenuated $T_{\rm eff}=1600 K$ model atmosphere for 2MASS 1207 b and the solid, blue curves are the atmosphere viewed through the additional dust shell.  Note that our 8.7$\micron$  upper-limit is drawn with a flat top at +1$\sigma$ and a downward arrow at the median value.  The models all assume 2.5 magnitudes of extinction at J-band, simulating the apparent under-luminosity of 2MASS 1207 b.  Shells at separations less than 20$R_{\rm b}$ are strongly ruled out by our 8.7$\micron$ +3$\sigma$ upper-limit.
\label{DUSTY}}
\end{figure}

The results of our shell models are presented in Figure \ref{DUSTY}.  Dust shells inside of 20$R_{\rm b}$ are ruled out by our Gemini/T-ReCS 8.7$\micron$ +3$\sigma$ upper-limit, while homogenous shells outside of this radius are not observed for objects of 2MASS 1207 b's age and mass.

\section{The Thick Cloud Atmosphere Hypothesis}\label{atmosphere}
JHK spectroscopy of 2MASS 1207 b shows its features to be consistent with a 1600 K AMES-DUSTY model atmosphere, but its luminosity is more consistent with a $\sim$1000 K model \citep{2007ApJ...657.1064M,2010AA...517A..76P}.  Typically, dust is thought to settle/condense below the photosphere for atmospheres $<$1200-1400 K \citep{2008ApJ...689.1327S}, so $\sim$1000 K brown-dwarfs/giant-planets are not expected to exhibit significant dust opacity.  However, models with cloud structures that are suspended above the photosphere in cooler atmospheres might be able to explain the spectral features of 2MASS 1207 b at the low temperatures indicated by its luminosity.  

Recently, \citet{2011arXiv1101.1973C} and \citet{2011arXiv1102.5089M} suggested that thick cloud atmospheres could explain the red colors and under-luminosity of the HR 8799 planets.  While their models do an adequate job reproducing the planets' photometry, a lack of detailed spectroscopic data diminishes the significance of their test.  In this section, we attempt to fit the available 2MASS 1207 b data (see Table \ref{published}) with the thick cloud atmospheres of \citet{2011arXiv1102.5089M}.  Medium-resolution, near-infared spectroscopy of 2MASS 1207 b \citep{2010AA...517A..76P} provides the strongest test yet of these models.

The \citet{2011arXiv1102.5089M} models \citep[see also][]{2006ApJ...640.1063B} parameterize cloud thickness from thickest (``Model-A", where the clouds extend to the top of the atmosphere), to thin (``Model-E") with two intermediate specifications (``Model-AE" and ``Model-AEE").  Model grids\footnote{available at \url{http://www.astro.princeton.edu/~burrows/8799/8799.html}} also vary $T_{\rm eff}$, $log(g)$, metallicity, cloud composition and modal grain size.  The A-models maintain a constant mixing ratio with decreasing pressure, and observationally are characterized by red near-infrared colors.  Similar colors are produced by the $f_{sed}=1$ models of \citet{2001ApJ...556..872A}, \citet{2008ApJ...689.1327S} and references therein, which might provide a similar explanation for 2MASS 1207 b's appearance.

\begin{figure}
 \includegraphics[angle=0,width=\columnwidth]{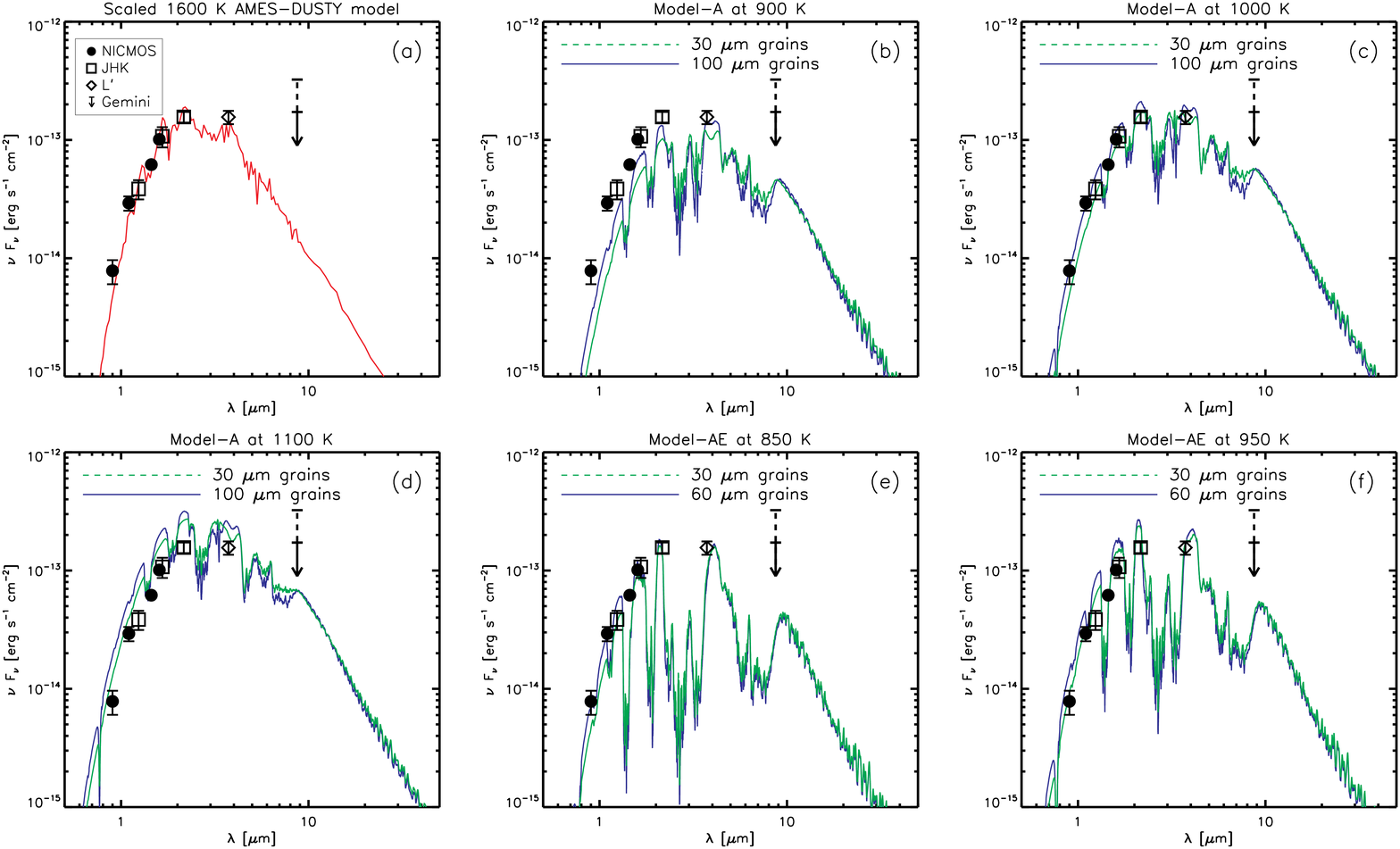}
\caption{Photometry of 2MASS 1207 b compared to (a) the best-fit, scaled AMES-DUSTY model \citep{2001ApJ...556..357A}, as described in \citet{2010AA...517A..76P}, (b)(c)(d) thick cloud models \citep[Model-A from][]{2011arXiv1102.5089M} at three different temperatures and with two different grain distributions, and (e)(f) intermediate (slightly thinner) cloud models \citep[Model-AE from][]{2011arXiv1102.5089M} at two different temperatures and with two different grain distributions.  The 1600 K AMES-DUSTY model is scaled to an unphysical object radius of 0.052 $R_{\Sun}$ (i.e. it is under-luminous or under-sized), whereas the cooler, \citet{2011arXiv1102.5089M} models assume an object size of $\sim$0.16 $R_{\Sun}$, based on the brown-dwarf/giant-planet cooling curves of \citet{1997ApJ...491..856B}.  If the \citet{2011arXiv1102.5089M} models are able to fully reproduce the photometry and spectroscopy of 2MASS 1207 b, without an unphysical radius scaling, they could explain the whole class of under-luminous brown-dwarfs/giant-planets.
\label{thickphotometry}}
\end{figure}

\begin{figure}
\begin{center}
 \includegraphics[angle=0,width=5.0in]{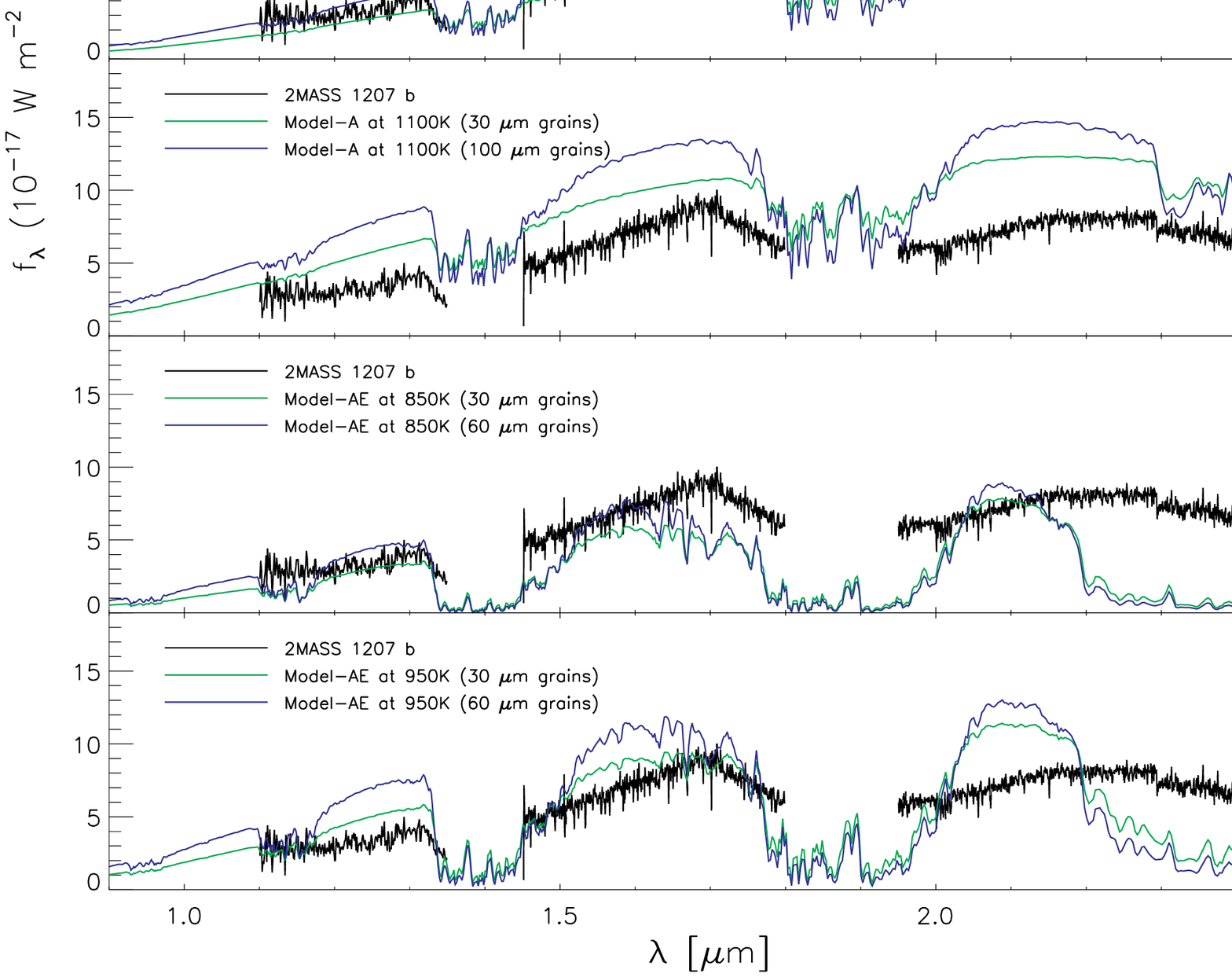}
\end{center}
\caption{The same models shown in Figure \ref{thickphotometry} compared to the JHK spectroscopy of \citet{2010AA...517A..76P}.  The 1600 K AMES-DUSTY model reproduces the spectral shape of 2MASS 1207 b, but is scaled to an unphysical object radius of 0.052 $R_{\Sun}$ (i.e. it is under-luminous or under-sized).  The cooler \citet{2011arXiv1102.5089M} models can explain 2MASS 1207 b's luminosity using a physically motivated object radius of $\sim$0.16 $R_{\Sun}$, but cannot explain its detailed spectral shape.  However, the \citet{2011arXiv1102.5089M} 1000 K A-models, shown in (c), are reasonable enough to suggest that modifications to the model might be able to explain 2MASS 1207 b's complete appearance.
\label{thickspectroscopy}}
\end{figure}

We plot a representative sample of the \citet{2011arXiv1102.5089M} models, as well as a best-fit, scaled AMES-DUSTY model \citep{2010AA...517A..76P} against the observed photometry (Figure \ref{thickphotometry}) and spectroscopy (Figure \ref{thickspectroscopy}) of 2MASS 1207 b.  In both figures, frame (a) shows the best-fit AMES-DUSTY model, frames (b), (c) and (d) show a sample of A-models (thick cloud), and frames (e) and (f) show a sample of AE-models (intermediate between thick and thin clouds).  All of the \citet{2011arXiv1102.5089M} models we show assume log(g)=4.0, solar metallicity, and  a forsterite cloud composition.

As described in \citet{2010AA...517A..76P}, the 1600 K, log(g)=4.5 AMES-DUSTY fits the spectral shape of 2MASS 1207 b adequately.  However, it can only fit the luminosity of 2MASS 1207 b by scaling the model radius to 0.052 $R_{\Sun}$, which is \textit{significantly} smaller than the $\sim$0.16 $R_{\Sun}$ radius predicted by evolutionary models for an object of 2MASS 1207 b's age \citep{1997ApJ...491..856B,2009AIPC.1094..102C}.  Both \citet{2011arXiv1102.5089M}'s A-models (thick cloud) and AE-models (intermediate) are able to approximately fit the overall SED of 2MASS 1207 b without an unphysical radius scaling \citep[the models simply assume a radius based on the evolutionary models of][]{1997ApJ...491..856B}.

The JHK spectroscopy of 2MASS 1207 b provides a stronger test of the \citet{2011arXiv1102.5089M} models than the SED photometry.  Of the AE models, the 850 K and 950 K models bound the measured luminosity of 2MASS 1207 b, but clearly do not fit the JHK spectral shape.  Of the A models, the 900 K and 1100 K models bound the measured luminosity of 2MASS 1207 b, while the 1000 K model gives the best fit.  The 1000 K A-model does not adequately fit the JHK spectroscopy of 2MASS 1207 b.  However, the fit is close enough, both in spectral shape, and magnitude so as to suggest that thick clouds might play a roll in explaining the under-luminosity of 2MASS 1207 b.  The 1000 K model with a modal grain size of 30$\micron$ fits the K-band spectrum of 2MASS 1207 b, although it overestimates the absorption of CO at 2.3 $\micron$ and does not fit the H-band spectrum.  The 1000 K model with a modal grain size of 100$\micron$ fits the H-band spectrum of 2MASS 1207 b, although it does not quite match the sharpness of the 2MASS 1207 b H-band spectrum and does not fit the K-band spectrum.

With a dedicated modeling effort, it might be possible to ``tweak" the thick-cloud 1000 K model to fit the spectroscopy of 2MASS 1207 b.  As shown in Figure \ref{thickspectroscopy}, frame (c), a modal grain-size of 30$\micron$ produces a model that is too red in H-K, and a modal grain-size of 100$\micron$ produces a model that is too blue.  Adjusting the grain-size distribution could address this problem.  Additionally, the sharpness of 2MASS 1207 b's H-band spectrum is not matched by the models.  Adjusting surface gravity might address this problem.  Finally, the models overestimate absorption at 2.3 $\micron$ due to CO.  Non-equilibrium chemistry, including the development of more complex organic molecules ($C_{2}H_{2}$, $C_{2}H_{4}$, etc.) as described in \citet{2009arXiv0911.0728Z}, might remove some of the carbon from CO and produce a better match of this feature.  Changing the parameterization of the cloud prescription could also help address some of these issues.

The adequate fit of the \citet{2010AA...517A..76P} spectra by the scaled 1600 K AMES-DUSTY model makes it tempting to choose this model as best describing 2MASS 1207 b.  However, we find that the 1000 K thick cloud atmosphere of \citet{2011arXiv1102.5089M}, with some modifications, may be able to fit 2MASS 1207 b's spectrum, without an unphysical radius scaling.  Using the cooling curve parameterizations of \citet{2001RvMP...73..719B} and 2MASS 1207 b's assumed age of 5-12 Myr, a $T_{\rm eff}=1000$K implies that 2MASS 1207 b's mass is between 5$M_{jup}$ and 7$M_{jup}$.

\section{Conclusions}

We have investigated three hypothetical explanations for the observed under-luminosity of 2MASS 1207 b: (1) the edge-on disk hypothesis proposed by \citet{2007ApJ...657.1064M}, (2) an isotropic dust-shell, motivated by the possible discovery of one around G 196-3 B \citep{2010ApJ...715.1408Z}, and (3) thick cloud model atmospheres \citep{2011arXiv1102.5089M} that might be able to fit simultaneously explain 2MASS 1207 b's spectral shape and luminosity.

We find the edge-on disk hypothesis unlikely for the following reasons:

1) Based on modeling and observations, young stars of all masses are expected to exhibit variability when occulted by an edge-on disk, as a result of the disk's non-axisymmetric structure and hydrostatic fluctuations.  Using data from HST/NICMOS \citep{2006ApJ...652..724S} and VLT/NACO (this work), we see no evidence for strong variability.  However, we caution that the magnitude and ubiquity of variability in edge-on brown dwarf disks are not currently well understood.

2) The inclination of 2MASS 1207 b has to be very tightly tuned to produce the observed under-luminosity effect.  At different inclinations, brown dwarfs can be \textit{un-extincted}, \textit{partially extincted}, or \textit{fully extincted} by their disks.  2MASS 1207 b falls into the regime of partially extincting disks, which our models predict only occur $\sim$6.4\% of the time in young brown dwarfs.  Since 2MASS 1207 b is such a unique system, this low probability is not, in itself, enough to completely rule out the edge-on disk hypothesis.

3) Since the discovery of 2MASS 1207 b and the ensuing discussion of its under-luminosity, several more systems have been found that appear to be under-luminous, including HD 203030 B, HN Peg B and HR 8799 bcde.  Since these other systems are likely all older than 2MASS 1207 b, their gas-rich disks will have dissipated, precluding the geometrically thick disk geometries necessary to extinct them.  Thus at least one other phenomenon must be capable of producing the same under-luminosity effect, rendering the edge-on disk hypothesis unnecessary.

While none of our three arguments individually rules out the edge-on disk hypothesis, collectively, they strongly suggest that other solutions to 2MASS 1207 b's apparent under-luminosity should be considered.  One possibility is that an isotropic dust-shell could provide the same extinction as an edge-on disk, without the geometric constraints that make an edge-on disk unlikely.  By enshrouding equilibrium model atmospheres with isotropic spheres of dust at different radii, we find the following:

1) Optically thick dust shells near the surface of the brown dwarf would emit blackbody radiation at high enough temperatures so that we would not observe ``gray-extinction" in the near-infrared.

2) Optically thick dust shells further from the surface (or at lower temperatures) would emit blackbody radiation in the mid-infrared.  Our new 8.7$\micron$ Gemini/T-ReCS photometry has a 1-$\sigma$ upper limit of 0.48 mJy and a 3-$\sigma$ upper limit of 0.92 mJy, which in our simplistic models, rules out dust shells at radii less than $\sim$20R$_{\rm b}$, or greater than $\sim$250 K.

Finally, we investigate the possibility that 2MASS 1207 b is not a $T_{\rm eff}=1600$ K object, despite the fact that its spectrum is well-fit by a 1600 K AMES-DUSTY model, scaled to a lower luminosity.  \citet{2011arXiv1101.1973C} have suggested that thick cloud models \citep[described in ][]{2011arXiv1102.5089M} can explain the photometric colors and overall luminosity of the HR 8799 planets, which, similar to 2MASS 1207 b, are under-luminous compared to AMES-DUSTY models.  We attempt to fit 2MASS 1207 b's photometry and spectroscopy with the \citet{2011arXiv1102.5089M} atmospheres and find the following:

1) The 1000 K A-models (thick clouds) are able to reproduce 2MASS 1207 b's low-luminosity.  The small ($a_{\rm mode}=30\micron$) dust-grain model mostly reproduce 2MASS 1207 b's K-band spectrum, although it overestimates CO absorption and does not reproduce the H-band spectrum.  The larger ($a_{\rm mode}=100\micron$) dust-grain model, mostly reproduces 2MASS 1207 b's H-band spectrum, although it under-estimates the sharpness of the H-band peak and does not reproduce the K-band spectrum.

2) Using $T_{\rm eff}=1000$K and assuming 2MASS 1207 b's age is between 5 and 12 Myr, we use the cooling curve scaling relations of \citet{2001RvMP...73..719B} to estimate that 2MASS 1207 b's mass is between 5$M_{jup}$ and 7$M_{jup}$.

With some ``tuning", it seems possible that the thick cloud models will be able to explain 2MASS 1207 b's JHK spectrum and photometry, without the need for luminosity scaling.  Currently, with its trigonometric distance, cluster-membership age and litany of spectroscopic and photometric data, 2MASS 1207 b provides the strongest test for the thick cloud models.  If their fit to the 2MASS 1207 b data is improved, it would signal that the under-luminous class of young brown-dwarfs/giant-planets are really cooler, lower-mass objects than their spectra imply.  It would also demonstrate that young brown-dwarfs/giant-planets are capable of suspending dust clouds in their photospheres at cooler temperatures than field brown dwarfs.

\acknowledgements
The authors thank the referee for his/her review, as well as Kevin Flaherty, Tom Greene, Mark Marley, Kevin Zahnle, Kees Dullemond and James Radomski for useful discussions.  We also thank Amanda Morrow and Kevin Luhman for providing us the IRS spectrum of 2MASS 1207 and Adam Burrows for supplying the thick-cloud model grid.  AJS acknowledges the NASA Graduate Student Research Program (GSRP) for its generous support.  LMC is supported by an NSF Career award and the NASA Origins of Solar Systems Program. LSz acknowledges support from the Spitzer Data Analyzes grant 1348621, the Hungarian OTKA grant K76816 and the Student Union of Science and Informatics of the University of Szeged.

\clearpage

\begin{deluxetable}{ccccccccccccc}
\tabletypesize{\scriptsize}
\tablecaption{Published Photometry/Spectroscopy of 2MASS 1207}
\tablewidth{0pt}
\tablehead{
\colhead{Reference} &
\colhead{Epoch} &
\colhead{Filter} &
\colhead{A Photometry (mag)} &
\colhead{b Photometry (mag)}
}

\startdata

\citet{2003tmc..book.....C} & May 1999        & J       & 13.00$\pm$0.03 &               \\
                            &                 & H       & 12.39$\pm$0.03 &               \\
                            &                 & Ks      & 11.95$\pm$0.03 &               \\
\hline
\citet{2003AJ....126.1515J} & Apr. 2002       & L'      & 11.38$\pm$0.10 &               \\
\hline
\citet{2004AA...427..245S}  & Jan. 2004       & 8.7$\micron$ & 5.6$\pm$1 mJy &           \\
                            &                 & 10.4$\micron$ & 7.5$\pm$1 mJy &          \\
\hline
\citet{2004AA...425L..29C}  & Apr. 2004       & H       &                & 18.09$\pm$0.21\\
                            &                 & Ks       &                & 16.93$\pm$0.11\\
                            &                 & L'      &                & 15.28$\pm$0.14\\
\hline 
\citet{2006ApJ...652..724S} & Aug. 2004       & F090M   & 14.66$\pm$0.03 & 22.34$\pm$0.35\\
                            &                 & F110M   & 13.44$\pm$0.03 & 20.61$\pm$0.15\\
                            &                 & F160W   & 12.60$\pm$0.03 & 18.24$\pm$0.02\\
                            & Apr. 2005       & F090M   & 14.71$\pm$0.04 & 22.58$\pm$0.35\\
                            &                 & F145M   & 13.09$\pm$0.03 & 19.05$\pm$0.03\\
                            &                 & F160W   & 12.63$\pm$0.02 & 18.27$\pm$0.02\\
\hline
\citet{2007ApJ...657.1064M} & Mar. 2005       & J       &                &  20.0$\pm$0.2 \\
                            & Apr.-Jun. 2005  & HK spectra & spectra     & spectra       \\
\hline
\citet{2006ApJ...639L..79R} & Jun. 2005       & IRAC3.6 & 8.49 mJy       &               \\
                            &                 & IRAC4.5 & 7.15 mJy       &               \\
                            &                 & IRAC5.8 & 6.36 mJy       &               \\
                            &                 & IRAC8   & 5.74 mJy       &               \\
                            &                 & MIPS24  & 4.32 mJy       &               \\
\hline
\citet{2008ApJ...676L.143M} & Jul. 2005\tablenotemark{a} & IRS spectra & spectra &             \\
\hline
\citet{2010AA...517A..76P}  & Jan.-Feb. 2007  & JHK spectra & spectra    & spectra       
\tablenotetext{a}{IRS spectra of 2MASS 1207 exist for Jul. 2005 and Jul. 2006. \citet{2008ApJ...676L.143M} published the 2005 data, but mistakenly listed the date as 2006 (Kevin Luhman, private communication, 2010).}

\enddata
\label{published}
\end{deluxetable}
\clearpage

\begin{deluxetable}{lcccccccccccc}
\tabletypesize{\scriptsize}
\tablecaption{T-ReCS 8.7 $\micron$ Observations of 2MASS 1207}
\tablewidth{0pt}
\tablehead{
\colhead{Date (UT)} &
\colhead{Used On-Source Time (s)} &
\colhead{IQ (\%)} &
\colhead{CC (\%)} &
\colhead{WV (\%)} &
\colhead{BG (\%)}
}

\startdata

Mar 29, 2008 & 304 & 70 & 50 & 50 & 50\\
Mar 31, 2010 & 4445 & 70 & 50 & 50 & Any\\
Apr 1, 2010 & 0 & 70/Any & 50 & 50 & Any

\enddata
\tablecomments{Gemini-South Observing Conditions (IQ=Image Quality, CC=Cloud Cover, WV=Water Vapor, BG=Background; descriptions available at \url{http://www.gemini.edu/sciops/telescopes-and-sites/observing-condition-constraints})}
\label{observations}
\end{deluxetable}

\begin{deluxetable}{lcccccccccccc}
\tabletypesize{\scriptsize}
\tablecaption{Archival NACO Ks photometry of 2MASS 1207}
\tablewidth{0pt}
\tablehead{
\colhead{Date (UT)} &
\colhead{Used On-Source Time (s)} &
\colhead{2MASS 1207 A Ks mag} &
\colhead{2MASS 1207 b Ks mag} &
\colhead{$\Delta$ mag b-A}
}

\startdata
2004 Apr. 27  & 480 & 12.07$\pm$0.02 & 16.96$\pm$0.06 & 4.88$\pm$0.05 \\
2005 Feb. 5  & 1560 & 12.16$\pm$0.02 & 16.77$\pm$0.03 & 4.61$\pm$0.03 \\
2005 Mar. 31 & 1560 & 12.15$\pm$0.02 & 16.73$\pm$0.03 & 4.58$\pm$0.03 \\
2009 Jun. 30 & 1860 & 12.23$\pm$0.10 & 17.20$\pm$0.22 & 4.97$\pm$0.20 \\

\enddata
\label{NACO observations}
\tablecomments{Absolute photometry is with respect to the nearby background star, 2MASS 12073400-3932586, which has a 2MASS Ks magnitude of 14.86 (and an error of 0.134, which we do not consider above).}
\end{deluxetable}

\clearpage

\clearpage

\clearpage

\clearpage

\clearpage

\clearpage

\clearpage

\clearpage

\clearpage

\clearpage

\bibliographystyle{apj}
\bibliography{database}

\end{document}